\pdfoutput=1
\RequirePackage{ifpdf}
\ifpdf 
\documentclass[pdftex]{sigma}
\else
\documentclass{sigma}
\fi

\begin{document}

\newcommand{\arXivNumber}{1410.7736}

\allowdisplaybreaks

\renewcommand{\PaperNumber}{006}

\FirstPageHeading

\ShortArticleName{Local Properties of Measures in Quantum Field Theory and Cosmology}

\ArticleName{Local Properties of Measures\\
in Quantum Field Theory and Cosmology}

\Author{Jos\'e M.~VELHINHO}

\AuthorNameForHeading{J.M.~Velhinho}

\Address{Faculdade de Ci\^encias, Universidade da Beira Interior,\\
R.~Marqu\^es D'\'Avila e Bolama, 6201-001 Covilh\~a, Portugal}
\Email{\href{mailto:jvelhi@ubi.pt}{jvelhi@ubi.pt}}

\ArticleDates{Received October 30, 2014, in f\/inal form January 14, 2015; Published online January 17, 2015}

\Abstract{We show that measure theoretical results concerning the Ashtekar--Lewandowski measure in the space of
generalized connections have direct analogues in the context of the Bohr compactif\/ication of the line and associated
Haar measure.
We present also a~cha\-racterization of the support of the measure associated with the canonical quantization of the free
massive scalar f\/ield, following closely well known analogous results concerning the Euclidean path integral measure.}

\Keywords{canonical quantization; scalar f\/ield; loop cosmology; support of the measure}

\Classification{28C10; 28C20; 81T99; 83F99}

\section{Introduction}

Measures in inf\/inite-dimensional spaces, linear and otherwise, feature prominently in quantum theory.
One can f\/ind them in the quantum theory of scalar f\/ields, both from the Euclidean path integral and from the canonical
quantization perspectives (see, e.g.,~\cite{GJ}), in the quantization of gravity (see~\cite{AL,T}) and in quantum
cosmology~\cite{ABL,AS}.
Typically, these measures are def\/ined on extensions of the spaces associated with the corresponding f\/ield theories or
classical mechanical systems.
In the quantum theory of scalar f\/ields one f\/inds measures in the space ${\mathcal S}'({\mathbb R}^d)$ of tempered
distributions, dual to the space of Schwartz functions.
In quantum gravity, in the so-called loop approach, one f\/inds measures in the space of generalized
connections~\cite{AL,MM,T,VS}, which is a~natural (though quite large) extension of the space of connections on
a~manifold.
In both cases the distributional extensions are crucial, given that, for measures of interest, the classical
non-distributional conf\/igurations corresponding to standard functions or connections turn out to be irrelevant sets,
i.e.~subsets of zero measure sets.
A~similar situation appears in the loop approach to quantum cosmology~\cite{ABL,AS,BCM,B}.
In this case a~role analogous to that of the space of generalized connections is played by the Bohr compactif\/ication of
the line, a~well know compact extension of~${\mathbb R}$.

In the present paper we characterize the support of two dif\/ferent measures of interest, in terms of local properties.
First, following closely~\cite{RR} and~\cite{CL}, we consider sets of measure zero and of measure one for the Gaussian
measure in ${\mathcal S}'({\mathbb R}^d)$ of covariance
\begin{gather*}
\langle \phi(f),\phi(g)\rangle = \frac{1}{2}\int d^dx\, f\big(m^2-\Delta\big)^{-1/2}g.
\end{gather*}
This is the measure uniquely associated with the free scalar f\/ield of mass~$m$, in~$d$-spatial dimensions, in the
canonical quantization formalism (see~\cite{GJ}), rather than the Euclidean path integral measure for~$d$-spacetime
dimensions considered in~\cite{CL, RR}.
Second, we show that measure theore\-ti\-cal results proved in~\cite{MTV}, concerning the so-called Ashtekar--Lewandowski
measure in the space of generalized connections, have close analogues in the context of the Bohr compactif\/ication of~${\mathbb R}$ and associated Haar measure.
In particular, we give a~local characterization of the support of this Haar measure and display an ergodic action which
in some sense mimics the action of spatial dif\/feomorphisms in the space of generalized connections.

\section{Canonical free f\/ield measure}
Consider the linear space of real Schwartz functions ${\mathcal S}({\mathbb R}^d)$, with its standard nuclear topology,
and let ${\mathcal S}'({\mathbb R}^d)$ denote its topological dual.
As is well known, a~scalar product $(\,,)$ on ${\mathcal S}({\mathbb R}^d)$, continuous with respect to the nuclear
topology, def\/ines a~Gaussian measure on ${\mathcal S}'({\mathbb R}^d)$, of mean zero and covariance
$\langle\phi(f),\phi(g)\rangle=(f,g)$.
(We will use the term {\em covariance} to denote also the scalar product $(\,,)$ that def\/ines the measure.) A~ref\/inement
of this result is given by Minlos' theorem for the case of Gaussian measures, which can be stated as follows
(see~\cite{RR,Ri,Ya}):

\begin{theorem}[Minlos]
\label{minlos}
Let~$E$ be a~real nuclear space, $(\,,)$ a~continuous scalar product and ${\mathcal H}$ the completion of~$E$ with respect to
$(\,,)$.
Let~$H$ be an injective Hilbert--Schmidt operator on ${\mathcal H}$, such that $E\subset H{\mathcal H}$ and $H^{-1}:E\to{\mathcal H}$
is a~continuous map.
Let $(\,,)_1$ be the scalar product on~$E$ defined by $(f,g)_1=(H^{-1}f,H^{-1}g)$. Then, the Gaussian measure on the dual
space $E'$ with covariance $(\,,)$ is supported on the subspace of $E'$ of those functionals which are continuous with
respect to the topology defined by $(\,,)_1$.
\end{theorem}

Let us mention also a~result by Colella and Lanford~\cite[Proposition~3.1]{CL}, applicable to Gaussian measures in
${\mathcal S}'({\mathbb R}^d)$ which are invariant under ${\mathbb R}^d$ translations.
Given a~covariance $(\,,)$ on ${\mathcal S}({\mathbb R}^d)$, its kernel ${\mathcal C}$ is the distribution on ${\mathbb
R}^{2d}$ def\/ined~by
\begin{gather*}
(f,g) =:\int d^dx\,d^dx' f(x){\mathcal C}(x,x')g(x'),
\qquad
\forall\, f,g\in{\mathcal S}\big({\mathbb R}^d\big).
\end{gather*}
Clearly, the corresponding measure in ${\mathcal S}'({\mathbb R}^d)$ is invariant under ${\mathbb R}^d$ translations
if and only if ${\mathcal C}(x,x')={\mathcal C}(x-x')$.
Let us recall that a~distribution $\phi\in {\mathcal S}'({\mathbb R}^d)$ can be identif\/ied with a~signed measure on
a~open set $U\subset{\mathbb R}^d$ if there is a~measure~$\nu$ on~$U$ and an integrable function~$F$ such that
$\phi(f)=\int_UfFd\nu$, for every $f\in{\mathcal S}({\mathbb R}^d)$ supported in~$U$.
Then~\cite{CL}:
\begin{theorem}
\label{rbprop1}
Let~$\mu$ be a~Gaussian measure in ${\mathcal S}'({\mathbb R}^d)$, invariant with respect to ${\mathbb R}^d$
translations and such that the kernel $\mathcal C$ of the corresponding covariance is not a~continuous function.
Then, for~$\mu$-almost every $\phi\in{\mathcal S}'({\mathbb R}^d)$ there is no non-empty open set $U\subset{\mathbb
R}^d$ on which~$\phi$ can be seen as a~signed measure.
\end{theorem}

\subsection{Local support properties}
Let us then consider the measure on ${\mathcal S}'({\mathbb R}^d)$ which provides the representation of the canonical
commutation relations associated with the free real scalar f\/ield of nonzero mass~$m$, in~$d$ spatial dimensions.
This is the Gaussian measure of covariance
\begin{gather}
\label{clivre0}
(f,g)_m:=\frac{1}{2}\int d^dx\,
f\big(m^2-\Delta\big)^{-1/2}g,
\qquad
f,g\in{\mathcal S}\big({\mathbb R}^d\big),
\end{gather}
where the mass is a~real parameter and~$\Delta$ is the Laplacian.
(See~\cite{GJ} for the relation between the Euclidean path integral, or constructive, formulation and the canonical
quantization approach.)

One can now apply the general results above to obtain both a~nontrivial zero measure set and nontrivial sets of measure
one, for the measure associated with the free f\/ield.

First, let us consider the kernel ${\mathcal C}_m$ of the covariance~\eqref{clivre0}.
It is clear that ${\mathcal C}_m(x,x')={\mathcal C}_m(x-x')$, and the distribution ${\mathcal C}_m(x)$ is easily obtained, after
Fourier transformation, as
\begin{gather*}
{\mathcal C}_m(x)=\frac{1}{(2\pi)^d}\int d^dp\,\frac{e^{ipx}}{2(m^2+p^2)^{1/2}}.
\end{gather*}
One \looseness=-1 can immediately conclude that, for any dimension~$d$ ($>$0) and any value of the mass~$m$, ${\mathcal C}_m(x)$ is not
a~continuous function, given that the Fourier transform $(m^2+p^2)^{-1/2}$ is not integrable, independently of~$d$.
So, by Theorem~\ref{rbprop1}, it follows that the set of distributions which, even if locally, can be identif\/ied with
signed measures, does not contribute to the measure def\/ined by~\eqref{clivre0}.

We will now apply Theorem~\ref{minlos} to the canonical free f\/ield measure def\/ined by~\eqref{clivre0}, in order to
obtain sets of measure one.
We follow here~\cite{CL,RR,Ri}, adapting the results therein to the canonical formalism.
Let us show that the distributions that support the measure are such that the action of the operator
$(1+x^2)^{-\alpha}(m^2-\Delta)^{-\beta}$ produces $L^2({\mathbb R}^d)$ elements, for $\alpha>d/4$ and
$\beta>(d-1)/4$.
To see this, let us consider the operators
\begin{gather*}
\widetilde H:=\big(m^2-\Delta\big)^{-\alpha}\big(1+x^2\big)^{-\alpha}
\end{gather*}
on $L^2({\mathbb R}^d)$, where $(1+x^2)$ is a~multiplication operator and $\alpha>d/4$.
Since $(1+x^2)^{-\alpha}$ is square integrable and the same is true for the Fourier transform $(m^2+p^2)^{-\alpha}$ of
$(m^2-\Delta)^{-\alpha}$, the operators $\widetilde H$ are of the Hilbert--Schmidt type $\forall\,\alpha >d/4$
(see~\cite[Lemma~1]{RR}).
Let us introduce the Hilbert space ${\mathcal H}^-_m$, which is the completion of ${\mathcal S}({\mathbb R}^d)$ with respect
to the inner product~\eqref{clivre0}.
Using the unitary transformation
\begin{gather*}
\big(m^2-\Delta\big)^{1/4}: \ L^2\big({\mathbb R}^d\big)\to{\mathcal H}^-_m,
\end{gather*}
we def\/ine the following Hilbert--Schmidt operator on ${\mathcal H}^-_m$:
\begin{gather*}
H=\big(m^2-\Delta\big)^{1/4}\widetilde H\big(m^2-\Delta\big)^{-1/4}.
\end{gather*}
Let us f\/inally introduce the scalar product in ${\mathcal S}({\mathbb R}^d)$:
\begin{gather}
(f,g)_1:=\int d^d x \big(\big(m^2-\Delta\big)^{-1/4}H^{-1}f\big) \big(\big(m^2-\Delta\big)^{-1/4}H^{-1}g\big)
\nonumber
\\
\phantom{(f,g)_1~}
=\int d^d x \big(\big(1+x^2\big)^{\alpha}\big(m^2-\Delta\big)^{\beta}f\big) \big(\big(1+x^2\big)^{\alpha}\big(m^2-\Delta\big)^{\beta}g\big),
\label{clivre1c}
\end{gather}
where $\beta=\alpha-1/4>(d-1)/4$.
By Theorem~\ref{minlos}, the subspace of those functionals which are continuous with respect to $(\,,)_1$ is a~set of
measure one.
Invoking the Riesz lemma, for each such functional~$\phi$ there is a~unique element~$f$ of the completion of ${\mathcal
S}({\mathbb R}^d)$ with respect to $(\,,)_1$ such that
\begin{gather*}
\phi(g)=(f,g)_1= \int d^d x \big(\big(m^2-\Delta\big)^{\beta}\big(1+x^2\big)^{\alpha}\big(1+x^2\big)^{\alpha}\big(m^2-\Delta\big)^{\beta}f\big) g.
\end{gather*}
Since $(1+x^2)^{\alpha}(m^2-\Delta)^{\beta}f$ above belongs to $L^2({\mathbb R}^d)$, one can say that continuous
functionals are uniquely identif\/ied as elements of the form $(m^2-\Delta)^{\beta}(1+x^2)^{\alpha}\psi$,
where $\psi\in L^2({\mathbb R}^d)$ and those elements are seen as functionals by means of integration (or the $L^2({\mathbb
R}^d)$ inner product).
The support of the measure can therefore be written as follows
\begin{gather}
\label{minlosup}
\big(m^2-\Delta\big)^{\beta}\big(1+x^2\big)^{\alpha} L^2\big({\mathbb R}^d\big),
\end{gather}
in the previously announced sense that the distributions which support the measure are such that the application of the
operator $(1+x^2)^{-\alpha} (m^2-\Delta)^{-\beta}$ produces elements of $L^2({\mathbb R}^d)$, for $\alpha>d/4$ and
$\beta>(d-1)/4$.

From a~local point of view, applying the operator $(m^2-\Delta)^{-\beta}$ to a~typical distribution is suf\/f\/icient to
produce a~locally $L^2$ function.
In other words, one can say that the Fourier transform of $(m^2+p^2)^{-\beta}\widetilde\phi(p)$ is locally $L^2$, for
almost every distribution $\phi(x)$, where $\widetilde\phi(p)$ denotes the Fourier transform.
In the most favourable case $d=1$, any negative power of $(m^2+p^2)$ achieves this ef\/fect.
Further applying the operator $(1+x^2)^{-\alpha}$ regularizes the behaviour at inf\/inity of typical distributions,
producing $L^2$ elements.

Let us conclude this section with a~couple of comments.

Note that although the mass~$m$ appears explicitly in the characterization of the support~\eqref{minlosup}, the space
that one obtains for support of the measure as a~consequence of Minlos' theorem is actually the same for all values of
the mass.
This can be seen, e.g.,
from the fact that the topology def\/ined by the scalar product~\eqref{clivre1c} is independent of the value of the mass
(as long as it is not zero).
So, the above characterization of the support is not sensitive to the value of the mass.
This can be traced back to the fact that $(m^2-\Delta)^{-1/2}({m'}^2-\Delta)^{1/2}$ is a~bounded operator, of bounded
inverse, $\forall\, m,m'$, and the situation should therefore not change with any other choice of Hilbert--Schmidt operator
in Theorem~\ref{minlos}.

Nevertheless, it is well known that the measures associated with two distinct values of the mass are in fact singular
with respect to each other.
Therefore, disjoint supports can be found, for distinct masses.
To unveil these crucial dif\/ferences in the support requires a~dif\/ferent type of analysis of the large scale behaviour of
typical distributions (see~\cite{MTV} for a~detailed study).

Since the dif\/ferences between the supports depend on the large scale, rather than on the local behaviour, the mutual
singularity of distinct free f\/ield measures disappears when the space ${\mathbb R}^d$ is replaced by a~compact
manifold.
From the measure theoretical perspective, this is at the root of the recent uniqueness of quantization results proved
in~\cite{3torus} and references therein.
In fact, mutual singularity of the measures translates into unitary inequivalence of the corresponding representations
of the canonical commutation relations, which is at the heart of the obstruction to generalize the results
of~\cite{3torus} to the noncompact case.

\section{Haar measure in the Bohr compactif\/ication of the line}
The second measure we wish to consider is the Haar measure on the Bohr compactif\/ication of the line.
The space in question, which we will denote by $\bar{{\mathbb R}}$, can be seen as the set of all group morphisms
from ${\mathbb R}$ to the unit circle~$T$, with an appropriate topology.

This space arises naturally in the approach to quantum cosmology known as loop quantum cosmology~\cite{ABL,AS,BCM,B,BS},
where it plays a~role analogous to that of the space of generalized connections in full blown loop quantum gravity.
It should be said, however, that from a~more physical viewpoint, the role of $\bar{{\mathbb R}}$ is less prominent
than that of the space of generalized connections.
In fact, although $\bar{{\mathbb R}}$ can be seen as the initial quantum conf\/iguration space for loop quantum
cosmology~\cite{ABL,bohr}, subsequent developments lead to an ef\/fectively smaller space.
The meaning of this is that whereas the quantum theory is initially constructed over the non-separable Hilbert space of
square integrable functions on $\bar{{\mathbb R}}$, one ends up with a~separable subspace thereof~\cite{AS, Vlqc}.
Also, in Bojowald's new quantization proposal~\cite{BS} one already starts from a~Hilbert space which is dif\/ferent from
the space of square integrable functions on $\bar{{\mathbb R}}$.

Let us then consider the line ${\mathbb R}$ with its commutative group structure given by addition of real numbers.
With the appropriate topology to be described below, the Bohr compactif\/ica\-tion~$\bar{{\mathbb R}}$ is the set ${\rm
Hom}[{\mathbb R},T]$ of all, not necessarily continuous, group morphisms from ${\mathbb R}$ to the multiplicative
group~$T$ of unitaries in the complex plane:
\begin{gather*}
\bar{{\mathbb R}}\equiv{\rm Hom}[{\mathbb R},T].
\end{gather*}
The generic element of $\bar{{\mathbb R}}$ is here denoted by $\bar x$.
So, every $\bar x\in\bar{{\mathbb R}}$ is a~map, $\bar x:{\mathbb R}\to T$ such that
\begin{gather*}
\bar x(0)=1
\qquad
\text{and}
\qquad
\bar x(k_1+k_2)=\bar x(k_1)\bar x(k_2),
\qquad
\forall\,
k_1,k_2\in{\mathbb R}.
\end{gather*}
Since~$T$ is commutative, it is clear that
\begin{gather}
\label{6}
\bar x  \bar x'(k):=\bar x(k)\bar x'(k)
\end{gather}
def\/ines a~group structure on $\bar{{\mathbb R}}$, which is again commutative.
Concerning the topology, note f\/irst that~$\bar{{\mathbb R}}$ is a~subgroup of the group of all (not necessarily morphisms) maps
from~${\mathbb R}$ to~$T$.
Since the latter can be identif\/ied with the product group~$\times_{k\in{\mathbb R}}T$, and~$T$ is compact, it carries
the Tychonof\/f product topology.
The product $\times_{k\in{\mathbb R}}T$ thus becomes a~compact (Hausdorf\/f) group.
This topology descends to~$\bar{{\mathbb R}}$, making it a~topological group.
Finally, from the fact that~$\bar{{\mathbb R}}$ contains only morphisms, one can easily check that it is a~closed subset of
$\times_{k\in{\mathbb R}}T$, and it is therefore compact.

So, $\bar{{\mathbb R}}$ is a~commutative compact group, with respect to the Tychonof\/f topology and the group operation~\eqref{6}.
Because it is compact, $\bar{{\mathbb R}}$ is equipped with a~normalized invariant (under the group operation) measure, namely
the Haar measure, here denoted by~$\mu_0$.

To proceed, it is convenient to introduce the projective structure of $\bar{{\mathbb R}}$, which we now review very brief\/ly
(see~\cite{bohr} for details).

For arbitrary $n\in{\mathbb N}$, let us consider f\/inite sets $\gamma=\{k_1,\ldots,k_n\}$ of independent real numbers
$k_1,\ldots,k_n$, with respect to the additive group operation in ${\mathbb R}$, i.e.\ such that the condition
\begin{gather*}
\sum\limits_{i=1}^n m_i k_i=0,
\qquad
m_i\in{\mathbb Z},
\end{gather*}
can only be fulf\/illed with $m_i=0$, $\forall\, i$.
(This notion of independence, used in~\cite{bohr}, is of course readily seen to be the same as linear independence over
${\mathbb Q}$.) Let $G_{\gamma}$ denote the subgroup of ${\mathbb R}$ freely generated by the set
$\gamma=\{k_1,\ldots,k_n\}$:
\begin{gather*}
G_\gamma:=\left\{\sum\limits_{i=1}^n m_i k_i,\; m_i\in{\mathbb Z}\right\}.
\end{gather*}
For each independent set~$\gamma$, let us introduce the group ${\mathbb R}_\gamma$ of all morphisms from $G_\gamma$
to~$T$,
\begin{gather*}
{\mathbb R}_\gamma:={\rm Hom}[G_\gamma,T].
\end{gather*}
It is shown in~\cite{bohr} that the family of spaces ${\mathbb R}_\gamma$ is a~well def\/ined projective family, and
that~$\bar{{\mathbb R}}$ is precisely the projective limit of this family.
Turning to measures, we know that measures in the projective limit~$\bar{{\mathbb R}}$ are in one to one correspondence with
(properly compatible) families of measures in the spaces of the projective family~\cite{Ya}.
So, the Haar measure in~$\bar{{\mathbb R}}$ is fully described by the family of measures that one obtains by push-forward, with
respect to the natural projections
\begin{gather}
\label{22}
p_\gamma: \ \bar{{\mathbb R}}\to{\mathbb R}_\gamma,
\qquad
\bar x \mapsto \bar x_{|\gamma},
\end{gather}
where $\bar x_{|\gamma}$ denotes the restriction of $\bar x$ to the subgroup $G_\gamma$.

Now, since each $G_\gamma$ is freely generated by an independent set~$\gamma$, each ${\mathbb R}_\gamma$ is
homeomorphic to a~$n$-torus $T^n$, with~$n$ being the cardinality of the set $\gamma=\{k_1,\ldots,k_n\}$.
Not surprisingly, one can easily check that the Haar measure on $\bar{{\mathbb R}}$ can be characterized as the (unique) measure
such that the push-forward of the projections~\eqref{22} gives precisely the Haar measure on the corresponding torus
$T^n$, $\forall$ independent set~$\gamma$.

\subsection{Support of the Haar measure}
We are now in position to give a~characterization of the support of the measure $\mu_0$, following in part arguments
presented in~\cite{MTV}.
\begin{proposition}
\label{teosupmal1}
The Haar measure $\mu_0$ on $\bar{{\mathbb R}}$ is supported on the set~$W$ of all elements $\bar x$ which are everywhere
discontinuous, as a~map from ${\mathbb R}$ to~$T$.
\end{proposition}

To prove this, it is suf\/f\/icient to show that the complement
\begin{gather*}
W^{\rm c}=\bigl\{\bar x\in \bar{{\mathbb R}}\,\big|\, \exists\, s_0\in {\mathbb R}\; \mbox{such that} \; \bar x\
\hbox{\rm is continuous in $s_0$}\bigr\}
\end{gather*}
is contained in a~zero measure subset of $\bar{{\mathbb R}}$.
Let us consider the sets
\begin{gather*}
\Theta_U=\bigl\{\bar x\in \bar{{\mathbb R}}\,\big|\, \exists\, I\; \mbox{such that}\; \bar x(I) \subset U\bigr\},
\end{gather*}
where $I\subset{\mathbb R}$ is an open set and~$U$ is a~measurable subset of~$T$ with $0<\mu_H(U)<1$.
Here, $\mu_H(U)$ is the Haar measure of~$U$.
We will need f\/irst the following

\begin{lemma}
\label{lemsupmal2}
The set $\Theta_U$ is contained in a~zero measure subset of $\bar{{\mathbb R}}$, for every $U\subset T$ such that $0<\mu_H(U)<1$.
\end{lemma}

To prove this lemma, let us start by recalling that the family of open balls
\begin{gather*}
B(q,1/m)=\bigl\{s\in [0,1]\,\big|\, |s-q|<1/m\bigr\}
\end{gather*}
with rational~$q$ and integer~$m$ constitutes a~countable basis for the topology of ${\mathbb R}$, and therefore
$\Theta_U$ is a~countable union of sets of the form
\begin{gather*}
\Theta_{U,q,m}:=\bigl\{\bar x\in \bar{{\mathbb R}}\,\big|\, \bar x\bigl(B(q,1/m)\bigr) \subset U\bigr\},\qquad
q\in {\mathbb Q},\qquad m\in{\mathbb N}.
\end{gather*}
One can now construct zero measure sets containing each of the $\Theta_{U,q,m}$ as follows.
For a~given pair $(q,m)$, let us choose an inf\/inite sequence $\{s_i\}_{i=1}^\infty \subset B(q,1/m)$ such that any of
the f\/inite subsets $\gamma_N:=\{s_1,\ldots,s_N\}$ is an independent set.
This is possible, since given any f\/inite independent set~$\gamma$ one can always f\/ind another point $x_0\in B(q,1/m)$
such that $\gamma\cup x_0$ is again independent.
Let us def\/ine the sets
\begin{gather*}
Z_N=\bigl\{\bar x\in \bar{{\mathbb R}}\,\big|\, \bar x(s_i)\in U,\; \forall\, s_i\in\gamma_N\},\qquad N\in{\mathbb N},
\end{gather*}
which satisfy $Z_{N+1}\subset Z_N$.
From the above mentioned relation between the measure $\mu_0$ and the measures obtained by means of the push-forward of
the map~\eqref{22}, it follows that
\begin{gather*}
\mu_0(Z_N)=\bigl(\mu_H(U)\bigr)^N,
\end{gather*}
leading to
\begin{gather*}
\mu_0\big(\cap_{N=1}^{\infty}Z_N\big)=\lim_{N\rightarrow\infty}\bigl(\mu_H(U)\bigr)^N=0,
\end{gather*}
by the~$\sigma$-additivity of the measure and $\mu_H(U)<1$.
On the other hand, $\Theta_{U,q,m}\subset Z_N$, $\forall\, N\in{\mathbb N}$, and therefore $\Theta_{U,q,m}$ is
contained in the set $\cap_{N=1}^{\infty}Z_N$, of zero measure.
It follows again from~$\sigma$-additivity of $\mu_0$ that, $\forall\, U \subset T$ with $\mu_H(U)<1$, the set $\Theta_U$
is contained in a~zero $\mu_0$-measure set.

Let us now conclude the proof of Proposition~\ref{teosupmal1}.
Let~$r$ be a~real number, $0<r<1$, and consider a~f\/inite cover of~$T$ by means of open sets $\{U_i\}_{i=1}^k$ such that
$\mu_H(U_i)=r$, which is clearly possible.
In fact, one can always take a~f\/inite subcovering of the covering consisting of neighbourhoods~-- of measure~$r$~-- of
each and every point in~$T$.
Let $\bar x$ be any given element of $W^{\rm c}$ and $s_0$ a~point of continuity.
Let $i_0$ be such that $\bar x(s_0)\in U_{i_0}$.
Continuity of $\bar x$ at $s_0$ implies that there exists a~neighbourhood $I\ni s_0$ such that $\bar x(I)\subset U_{i_0}$.
Therefore, every element of $W^{\rm c}$ belongs to one of the sets $\Theta_{U_i}$, i.e.~$W^{\rm c}\subset\cup_{i=1}^k\Theta_{U_i}$.
This concludes the prove of the Proposition, given Lemma~\ref{lemsupmal2}.

\subsection{Ergodic action}

Let us consider the multiplicative action of ${\mathbb R}\backslash \{0\}$ on ${\mathbb R}$,
$(\lambda,k)\mapsto\lambda k$, which obviously preserves the additive group structure, i.e.~$\lambda(k_1+k_2)=\lambda k_1+\lambda k_2$.
We therefore also have an action of~${\mathbb R}\backslash \{0\}$ on~$\bar{{\mathbb R}}$, $(\lambda,\bar x)\mapsto\lambda^* \bar x$, such that
\begin{gather}
\label{action}
\lambda^* \bar x(k):=\bar x(\lambda k).
\end{gather}
It is straightforward to check that the Haar measure $\mu_0$ is invariant under this action.
By standard arguments (see~\cite{AL,T}), it is suf\/f\/icient to consider the integration of so-called cylindrical
functions.
In the present case these are essentially complex functions on a~given~$n$-torus, since they are def\/ined as compositions
of complex functions on the sets ${\mathbb R}_\gamma$ with the projections $p_\gamma$~\eqref{22}.
So, a~general cylindrical function is of the form
\begin{gather*}
F(\bar x)=f\left(\bar x(k_1),\ldots,\bar x(k_n)\right),
\end{gather*}
where $\gamma=\{k_1,\ldots,k_n\}$ is an independent set and $f:T^n\to {\mathbb C}$ is an integrable function.
Under the above action~\eqref{action}, such a~cylindrical function transforms into
\begin{gather*}
\lambda^*F(\bar x)=f\left(\bar x(\lambda k_1),\ldots,\bar x(\lambda k_n)\right).
\end{gather*}
So, the transformation amounts to switching from an independent set~$\gamma$ to another independent set
$\lambda\gamma:=\{\lambda k_1,\ldots,\lambda k_n\}$, of the same cardinality.
Now
\begin{gather*}
\int_{\bar{{\mathbb R}}}Fd\mu_0=\int_{T^n}f\,d({p_{\gamma}}_*\mu_0)
\qquad \text{and}\qquad
\int_{\bar{{\mathbb R}}}\lambda^*Fd\mu_0=\int_{T^n}f\,d({p_{\lambda\gamma}}_*\mu_0),
\end{gather*}
where ${p_{\gamma}}_*\mu_0$ and ${p_{\lambda\gamma}}_*\mu_0$ are given by push-forward of $\mu_0$ with respect to
$p_\gamma$ and $p_{\lambda\gamma}$.
(In fact, we are not showing explicitly the maps relating ${\mathbb R}_\gamma$ and ${\mathbb R}_{\lambda\gamma}$
to $T^n$.
The argument is nevertheless rigorous, see, e.g.,~\cite{AL,T} for totally similar arguments concerning the
Ashtekar--Lewandowski measure.) But the measures ${p_{\gamma}}_*\mu_0$ and ${p_{\lambda\gamma}}_*\mu_0$ are actually the
same, both being equal to the Haar measure on $T^n$, and we get
\begin{gather*}
\int_{\bar{{\mathbb R}}}F\,d\mu_0=\int_{\bar{{\mathbb R}}}\lambda^*F\,d\mu_0.
\end{gather*}
Since this is true for every cylindrical function, one can conclude that the measure $\mu_0$ is invariant.
\begin{proposition}
The Haar measure on~$\bar{{\mathbb R}}$ is ergodic with respect to the action of ${\mathbb R}\backslash \{0\}$~\eqref{action}.
\end{proposition}

One can prove this result following lines of reasoning previously presented in~\cite{MTV}.
We will use the well known fact that a~measure is ergodic with respect to a~group action if and only if the constant
functions are the only (measurable) functions that remain invariant under the action of the group (see, e.g.,~\cite{Ya}).
In the present case   this amounts to show that the only elements of~$L^2(\bar{{\mathbb R}},\mu_0)$ that remain invariant under
the action~\eqref{action} are the constant functions.

Let us f\/irst remind that the set of functions $\{F_k, k \in{\mathbb R}\}$ def\/ined by
$
F_k(\bar x)=\bar x(k)
$
forms a~complete orthonormal set on $L^2(\bar{{\mathbb R}},\mu_0)$ (see, e.g.,~\cite{bohr}).
It is clear that under the action (induced from)~\eqref{action}, the functions $F_k$ transform amongst themselves:
\begin{gather}
\label{8}
F_k\mapsto F_{\lambda k}.
\end{gather}
Let then~$\psi$ be an element of $L^2(\bar{{\mathbb R}},\mu_0)$.
Since the set $\{F_k, k \in{\mathbb R}\}$ is an orthonormal basis, one can write~$\psi$ in the form
\begin{gather}
\label{9}
\psi=\sum\limits_{k \in{\mathbb R}}c_kF_k,
\end{gather}
with no more than countably many coef\/f\/icients $c_k$ being nonzero.
Given the transformation~\eqref{8}, one concludes that~$\psi$ is invariant under the action~\eqref{action} if and only
if
\begin{gather}
\label{10}
c_k=c_{\lambda k},
\qquad
\forall\, k\in{\mathbb R},
\qquad
\forall\, \lambda\in{\mathbb R}.
\end{gather}
Suppose there is $k\not =0$ such that $c_k\not = 0$.
Then it follows from~\eqref{10} that in the sum~\eqref{9} there is an inf\/inite number of equal~-- nonzero~-- coef\/f\/icients,
which is clearly impossible.
One then concludes that an invariant~$\psi$ is proportional to the identity function $F_0$, i.e.~is a~constant.

\subsection*{Acknowledgements}
I would like to thank Jos\'e Mour\~ao and also the anonymous referees for comments and corrections.

\pdfbookmark[1]{References}{ref}
\LastPageEnding

\end{document}